\documentclass[preprint2]{aastex63}
\usepackage{amssymb }

\accepted{21 Dec 2021}

\begin{document}

\title{The mid-UV spectrum of irradiated NaCl at Europa-like conditions}

\correspondingauthor{M.E. Brown}
\email{mbrown@caltech.edu}

\author[0000-0002-8255-0545]{Michael E. Brown}
\affiliation{Division of Geological and Planetary Sciences\\
California Institute of Technology\\
Pasadena, CA 91125, USA}
\author{William T.P. Denman}
\affiliation{Division of Geological and Planetary Sciences\\
California Institute of Technology\\
Pasadena, CA 91125, USA}
\author{Samantha K. Trumbo}
\affiliation{Division of Geological and Planetary Sciences\\
California Institute of Technology\\
Pasadena, CA 91125, USA}
\begin{abstract}
Recent observations from the {\it Hubble Space Telescope} show a mid-UV absorption
feature localized to leading hemisphere chaos regions on Europa. The same
regions were previously found to have a visible absorption at 450 nm that
was attributed to the presence of irradiated NaCl. The lack of any
additional diagnostic absorptions for NaCl in the visible spectrum of these terrains
made confirmation of
this identification difficult. Here we use laboratory experiments to show that NaCl irradiated
at Europa's surface temperatures develops an absorption at $\sim$220 nm consistent
with the new detection in Europa's mid-UV spectrum,  strongly supporting the NaCl identification. Irradiated NaCl in leading- hemisphere chaos terrain would suggest
that sodium and chlorine are
important components of Europa's subsurface ocean.
\end{abstract}


\section{Introduction} 
Jupiter's satellite Europa has an icy surface that shows
spectroscopic evidence for the presence of salts, 
particularly in the regions
that appear to have been most recently geologically resurfaced
\citep{1999JGR...10411827M}.
These salts are likely ultimately derived from the
large interior ocean of Europa and thus record information on
the composition of this ocean. Most of the information on
the composition of these salts has come from near-infrared 
spectroscopy. Unfortunately, the near-infrared spectra of
Europa are ambiguous: distorted water ice bands indicative of
hydrated minerals are clearly present, but few diagnostic
spectral bands are present that would allow unique identification 
of the salts.

Owing to these difficulties, three major interpretations for
the native salt composition of the surface of Europa have 
been proposed.  Galileo NIMS data were largely interpretted
as showing the presence of magnesium sulfate salts as 
well as radiolytically produced sulfuric acid on the sulfur-bombarded
trailing hemisphere \citep{1998Sci...280.1242M,1999Sci...286...97C, 2005Icar..177..472D,2010Icar..210..358S,2012JGRE..117.3003D}. Ground-based observations from the Keck
Observatory at higher 
spectral resolution confirmed the presence of sulfuric
acid on the trailing hemisphere, but
failed to detect any of the distinctive spectral
features expected from magnesium sulfates on most of the surface, leading
to the hypothesis of a native salt composition dominated
by spectrally-bland sodium chloride, with the sodium chloride
domnating leading hemisphere chaos regions
\citep{2013AJ....145..110B,2015AJ....150..164F}. 
Though   subsequent observations
from the Very Large Telescope did not detect any additional
spectral features, they were modeled as a combination
of (presumably native) magnesium chloride salt as well
as (presumbably radiolytic) magnesium chlorates, magnesium
perchlorates, and
sulfuric acid \citep{2016AJ....151..163L}, with the chloride
salts dominating on the trailing, rather than leading side in this
model.
With few spectral features in the near-infrared
on which to base these interpretations, resolution of this
compositional ambiguity is difficult.

Pure chloride salts tend to be spectrally featureless in the visible
wavelength region, but irradiation can change these characteristics.
Irradiated sodium chloride, in particular, grows spectral
absorption features owing to the presence of electrons 
occupying lattice locations vacated by chlorine ions \citep{RevModPhys.18.384}.
\citet{2015GeoRL..42.3174H} and \citet{2017JGRE..122.2644P} studied these visible absorptions
-- called ``color centers'' -- at Europa-like conditions
and showed the presence of a broad 460 nm absorption (called
an ``F-center'') due to these electrons as well as an 
absorption at 720 nm (an "M-center") due to physically
adjacent pairs of F-centers. Observations from the
Hubble Space Telescope (HST) showed a distinct
absorption feature at 450 nm, interpretted to be F-center
absorption, but no hint of the expected
M-center absorption at 720 nm. \citet{2019SciA....5.7123T} hypothesized
that a balance between radiolytical F-center production
 and photobleaching -- the destruction of F-centers by 
photo-excitation of the trapped electrons -- led to
a low equilibrium number of F-centers too small to ever
have sufficient numbers of adjacent pairs to 
begin to form M-centers. This hypothesis was confirmed
by \citet{denman_nacl_2022}, who showed that
more carefully 
controlled temperatures led to F-centers shifted to the HST-observed
450 nm wavelengths and that solar-like
illumination conditions can lead to the
presence of an F-center with no M-centers, 
matching the Europa observations.

Recently, \citet{trumbo_HSTUV_2022} detected a second
spectral absorption  in the same leading-hemisphere
chaos regions where the absorption attributed to the NaCl
F-center was detected.
This new mid-ultraviolet (UV) feature,
with a central wavelength of approximately 230 nm,
occurs near the reported wavelength of the V$_3$
color center in NaCl, a feature that occurs due the displacement
of Cl atoms from their normal lattice positions and 
the subsequent formation of 
Cl$_3^{-1}$ molecular ions in nearby 
Cl lattice sites \citep{V3_1949,V3,PhysRevB.78.024120}.
V$_3$ centers
are stable up to $\sim$ 500K and have a poorly determined 
central absorption wavelength near 210~nm. 
The UV spectra could also be affected
by the presence of H-centers with an absorption near 330~nm, 
which are caused by Cl$_2^{-1}$
molecular locations in Cl lattice sites but which are
thought to be unstable at temperatures above about 80 K \citep{PhysRevB.78.024120}.
These features have never been studied at the temperatures
of the surface of Europa nor in reflection spectroscopy, 
yielding comparison to the HST data difficult.
Here we measure the mid-UV spectrum of irradiated NaCl 
at Europa-like conditions from 200 to 700~nm to allow for
direct comparison to the HST spectra of \citet{trumbo_HSTUV_2022}.

\section{Experimental setup}
The experimental setup for observing irradiated 
NaCl is described in detail in \citet{denman_nacl_2022}. In short,
we irradiate  $\sim$500 $\mu$m NaCl grains with 10 keV electrons in a 
16 cm diameter vacuum chamber kept at a pressure of
$~1\times 10^{-8}$ torr and at controllable
temperatures. To ensure thermal coupling between
the salt grains and the cold finger, the grains are pressed
into indium foil inside of the sample holder
and any excess
grains are poured off. The temperatures we report are
of the cold finger itself, so even with the indium foil the sample 
temperature could still be somewhat elevated above this temperature. 
Because of this procedure,
small amounts of indium are visible in the sample cup, but our calibration
ensures that the visibility of indium does not affect the 
final results, and we have verified that irradiating indium
with electrons at the same dosage level used here
does not change its mid-UV spectrum by more than $\sim$2\%.

To obtain spectra, we illuminate the sample with a 
stabilized deuterium UV lamp
at a 45$^\circ$ angle to the sample, and, to obtain a
diffuse reflection, we observe at an angle 90$^{\circ}$ away
from the specular beam. The beam illuminates only the sample
inside the holder, ensuring that all observed light comes from 
the sample. The diffusely reflected beam is fed into
a fiber-coupled spectrometer which
obtains a 200 to 1100 nm spectrum in a single exposure. 
To maximize the signal-to-noise down to 200 nm, where 
little light from the lamp is created, we collect long exposures
and allow the spectrum to
saturate in some lamp emission features redward of 500 nm.
In the data analysis, we extrapolate over these saturated regions.

Within the chamber, relative reflectance calibration is obtained 
with respect to the unirradiated NaCl.
We calibrate the absolute reflectance of the NaCl 
exterior to the chamber by comparing to the 
spectrum of PTFE powder freshly pressed to a density of
1.025 g cm$^{-3}$, which is is essentially
uniform except for a $\sim$2\% decline 
in reflectance from 500 nm to 200 nm \citep{PTFE}.
Comparison of the PTFE to a full NaCl sample (with no indium
foil showing) shows that NaCl is nearly uniformly reflective
across the UV-visible, but that a small absorption begins blueward
of 240 nm. Experimentation with grinding the NaCl grains to 
smaller sizes shows that these features become smaller and
should be essentially unobservable at the tens of microns
sizes of the particles thought to be on Europa \citep{1998Sci...280.1242M}. For the remainder
of our analysis we will therefore assume that unirradiated
NaCl is spectrally uniform across our wavelength range,
and we will reference all of our spectra to that of
the cold unirradiated
NaCl in vacuum. 

The spectrum of NaCl Within the chamber will contain small amounts
of contamination from the
spectrum of indium, which is marginally visible between NaCl grains. As we are referencing
to the unirradiated sample, the spectral contamination will be removed
by the reference, unless the indium changes when irradiated. To ensure that 
the indium does not influence the spectral results, 
we irradiated pure indium foil at the full dosage of the experiments described
below. No spectral changes were found at the $\sim$1.5\% 
level { (see Appendix)}. In addition, we performed room temperature NaCl irradiation
experiments with no indium, and, while the growth rate and central
wavelength of the spectral features
change due to the higher temperatures, 
all of the spectral features seen in the experiments described
below with indium are
reproduced. We are thus confident that the spectral changes seen are due to changes
in the NaCl.

We perform two series of experiments, bracketing the
expected 120 K dayside and 80 K nightside temperature of the Europa equator. 
In both experiments
we irradiate the grains with a current density of
12 nA cm$^{-2}$. Assuming a 1.2 $\mu$m penetration 
depth for these 10 keV electrons \citep{2017JGRE..122.2644P}, this
current density provides $\sim$250 times the energy flux
as received at the top 1.2 $\mu$m of the surface of Europa. 
\citet{denman_nacl_2022} showed the importance of photobleaching 
during irradiation in reaching an equilibrium between
F-center creation and destruction.
Our UV lamp provides little illumination at the 450 nm wavelength
of the F-center, so, when not obtaining spectra, we switch on
the visible spectral lamp, which provides approximately
13 times less than the noon-time equatorial solar irradiation
at Europa. Because of the mismatch between the
amount of irradiation and the amount of photobleaching,
we do not reach equilibrium and F-centers continue to
grow. We thus terminate our experiments when the features
are fully developed but before saturation. 

\section{Results}
The results of irradiation at 80 K and at 120 K are shown
in Fig 1. In both cases, a well-defined
F-center at 450 nm develops within the first 15 minutes,
and a weaker band in the mid-UV begins to appear.
At our low dose levels M-centers at 720 nm never appear.
Small differences
between the 80~K and 120~K irradiation are apparent.
The 80~K irradiation develops a broad band 
from 300-400~nm
on the shoulder of the F-center band that is not apparent
in the 120~K radiation, and at 80~K the 
F-center grows more slowly and is slightly shifted
blueward. The mid-UV band
of the 80~K irradiation is initially
shifted redward with respect to 
120~K, but by the end of the irradiation the mid-UV feature
at 80~K and 120~K appears identical. 
\begin{figure}
\epsscale{1.2}
\centering
\plotone{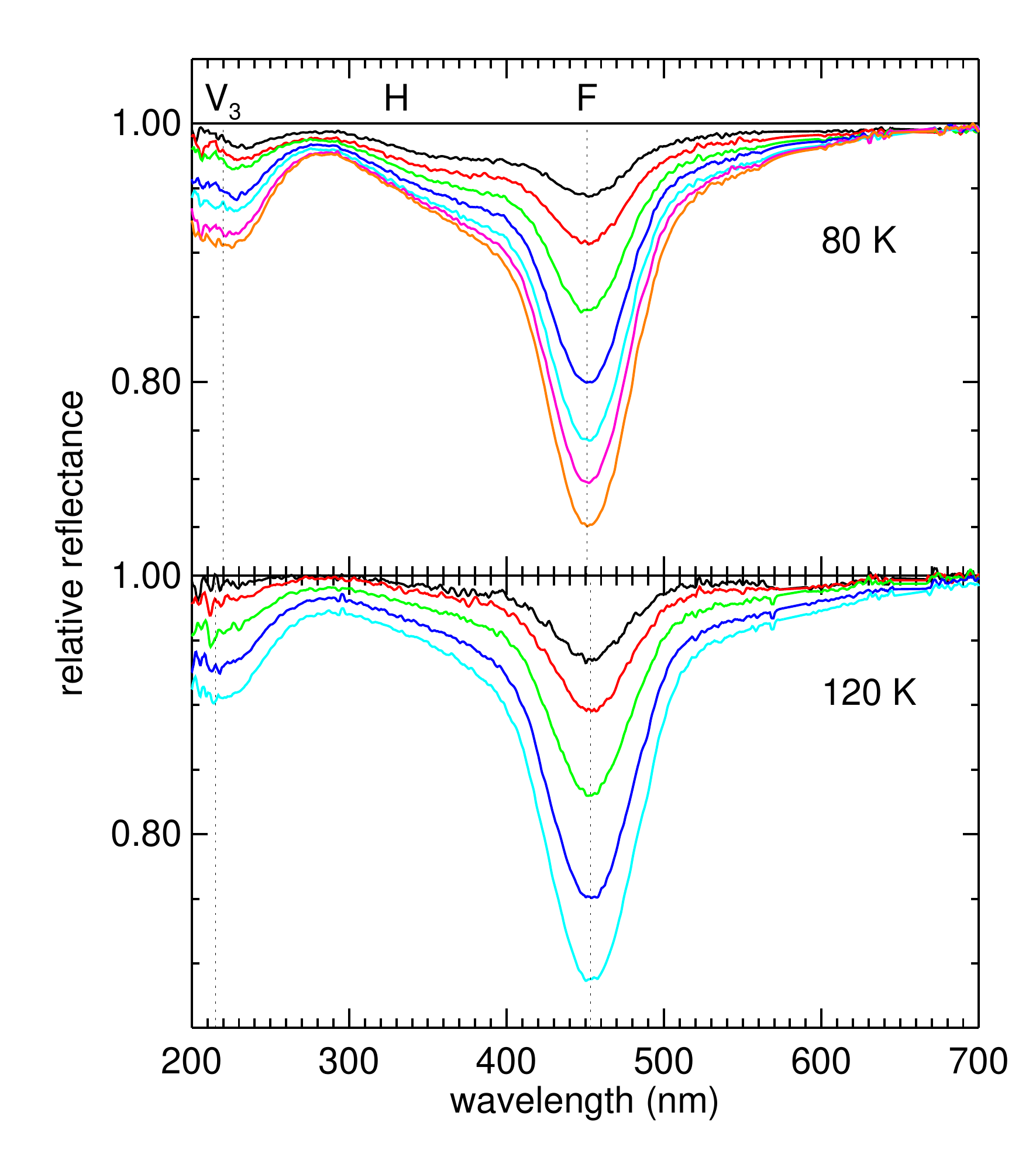}

    \caption{The spectrum of NaCl irradiated at 80 and 120 K with
    10 keV electrons. Spectra are shown relative to the spectrum
    of unirradiated NaCl, which is expected to be nearly spectrally
    flat through this region. The irradiation times are
    15 minutes (black), 30 minutes (red), 1 hour (green),
    2 hours (blue), 3 hours (tourquoise), 4 hours (magenta),
    and 5 hours (orange). For the 80 K irradiation, dashed lines at 220 and 451 nm
    show the central wavelengths of the color
    centers, while at 120 K these shift to 453 and 215 nm.
    }
\end{figure}
The absorption from 300-400 nm is likely due to
the presence of H-centers in the colder sample. These defects
become unstable at temperatures above $\sim$80~K and recombine with F-center defects \citep{PhysRevB.78.024120}. 
The change in the F-center wavelength with 
temperature has been previous noted \citep{PhysRevB.78.024120,denman_nacl_2022}.

To further explore the effects of temperature and of photobleaching,
we terminated the electron irradiation and warmed the~80 K
sample to 120~K. After 2 hours of photobleaching 
the UV feature remains unchanged while the 
F-center has shifted redward and begun to decay. 
The H-center absorption appears to be gone completely.

\begin{figure}
\epsscale{1.2}
\plotone{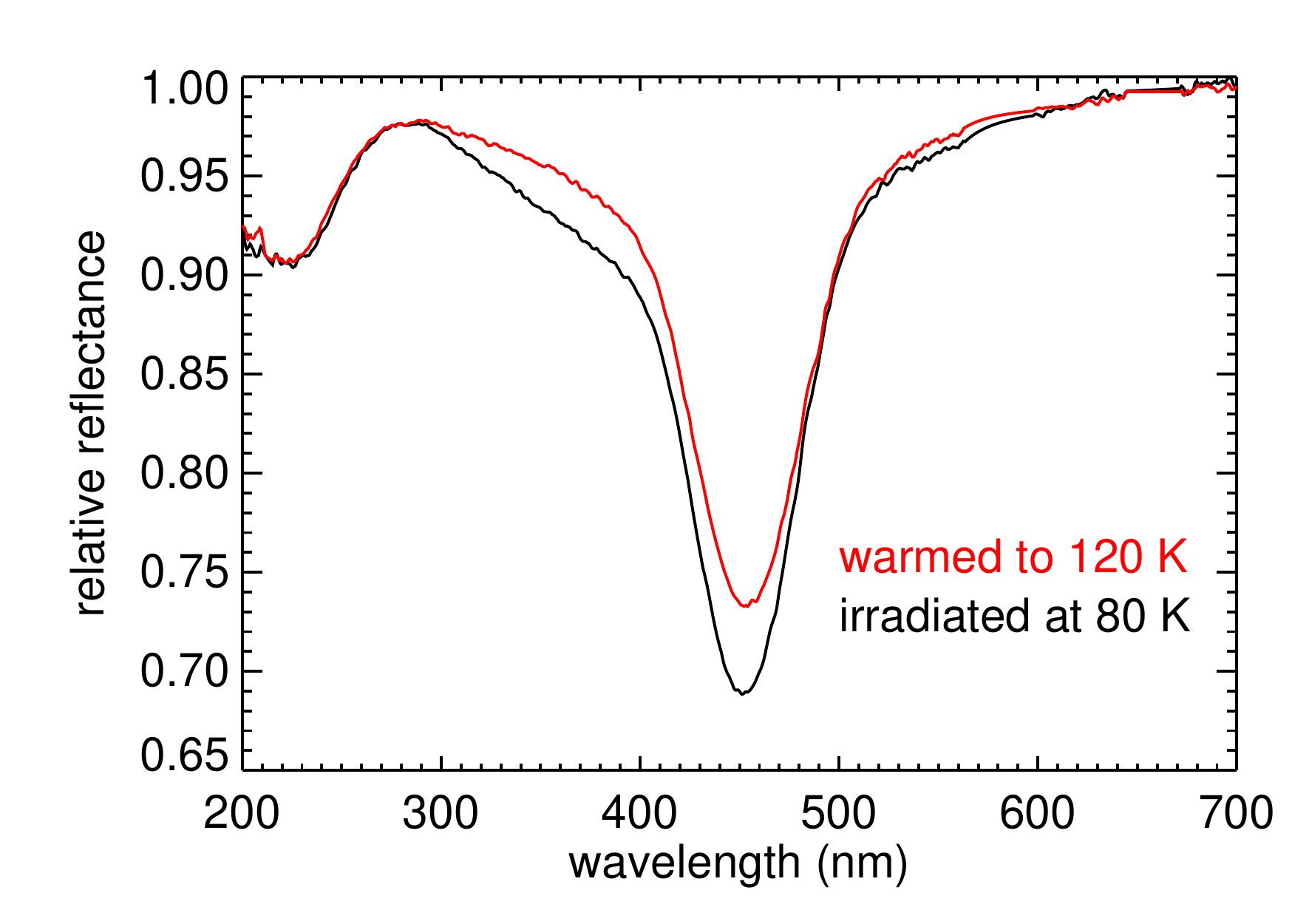}
    \caption{NaCl irradiated at 80 K for 5 hours (black) 
    compared to the same sample after termination of radiation,
    raising of the temperature to 120~K, and photobleaching at
    Europa-like fluxes for 2 hours. The F-center absorption decays and 
    shifts redward with higher temperatures and photobleaching,
    the H-center absorption appears to dissipate entire,
    and the V$_3$ absorption remains unchanged.}
\end{figure}

Given that the ratio of photobleaching to irradiation experienced
by Europa is significantly higher than in our experiments, it
is plausible that this last spectrum could best represent the
surface conditions of Europa, with 80~K irradiation at night
allowing color centers to grow and then~120 K temperatures
with intense
photobleaching during the day to causing them to decay back away.
The difference in decay rate between the photobleached V$_3$ and F- centers and the 
likely difference between the photon and electron penetration depth suggests that at equilibrium the ratio of 
the features could differ from that seen in our experiments. 

\section{Implications for Europa}
NaCl irradiated at the temperatures of the surface of Europa
develops an absorption band from $<200$~nm to 280~nm
which is broadly consistent with the feature detected
from HST spectroscopy on Europa \citep{trumbo_HSTUV_2022}. 
This V$_3$ color center and the F color center at 450~nm
are the only distinct absorption features 
seen in our experiments.
The HST results show that the UV absorption
and the 450~nm absorption are spatially coincident, localized 
to leading hemisphere chaos regions. 
These results, along with the photobleaching experiments
showing that M-center absorption at 720~nm should not
be expected on Europa \citep{denman_nacl_2022}, support a consistent conclusion that
irradiated NaCl is present in the leading-hemisphere chaos
terrains of Europa.

NaCl on Europa is unlikely to be exogenic. Though NaCl is
present in the Jupiter system owing to a source on Io \citep{2003Natur.421...45L},
the spatial segregation of the NaCl to specific geographic
regions is difficult to explain unless the NaCl is endogenic
to Europa. Geologically young chaos terrains are a natural
location to find salts recently sourced from the interior. 
While the chaos terrains may not directly sample ocean water,
their endogenic composition must nonetheless be
controlled by subsurface chemistry.

In the simple slow-freezing framework and experiments of
\citet{2019Icar..321..857J}, the presence of NaCl suggests
the freezing of a sodium and chlorine rich, but relatively sulfate poor, brine.
In such a case abundant sulfate salts on the surface would
come from sulfur bombardment and would presumably be 
concentrated on the trailing hemisphere similarly to the
radiolytically produced 
sulfuric acid \citep{1999Sci...286...97C, 2015AJ....150..164F}.
MgCl$_2$ salts, as proposed by \citep{2016AJ....151..163L}, could
also freeze from such brines if the Mg content is sufficiently
high. Radiolytic or photolytic processing could then yield 
perchlorides and perchlorates of magnesium
\citet{2019Icar..321..857J}, though their proposed spatial
distribution on Europa would be difficult to explain. The attempt to determine
the surface composition of Europa using linear spectral
modeling of \citet{2016AJ....151..163L} ruled out NaCl
in favor of these Mg-bearing chlorinated salts. The detection
of two distinct spectral features consistent with those expected from
irradiated NaCl
perhaps highlights the difficulties
of using such modeling to infer detailed surface 
composition. With distinct spectral features expected from
NaCl now firmly detected, 
reliably determining  the presence or absence of chlorinated
magnesium salts and of sulfates would be a major step
forward in understanding the chemistry of the ocean of 
Europa and the radiolytic processing that occurs on its surface.

\acknowledgements
This research was supported by a grant number 668346
from the Simons Foundation.

\section{Appendix}
{
Sodium chloride crystals pressed into indium foil provide significantly better thermal
coupling than when the crystals are simply sitting on a cold finger
\citep{denman_nacl_2022}.
In our experiments, we press the crystals into the foil and pour off any crystals which are 
not sufficiently embedded into the foil to stick. Such a process leaves a sample which is
nearly completely covered in crystals, but through which the indium can be see through a small
fraction of the surface area (Figure A1). To ensure that the changes that were observed in
our sample upon irradiation were due solely to effects in the sodium chloride, we
irradiated bare indium to search for any radiation-induced changes. Reproducing the same
flux and dosage as the sodium chloride experiments, we find changes of no more than $\pm 1.5$\% in the indium (Figure A2).
These very small changes, coupled with the small fraction of indium visible, ensure that the 
all changes seen in the spectra are due to sodium chloride.}

\begin{figure}
    \centering
	\plotone{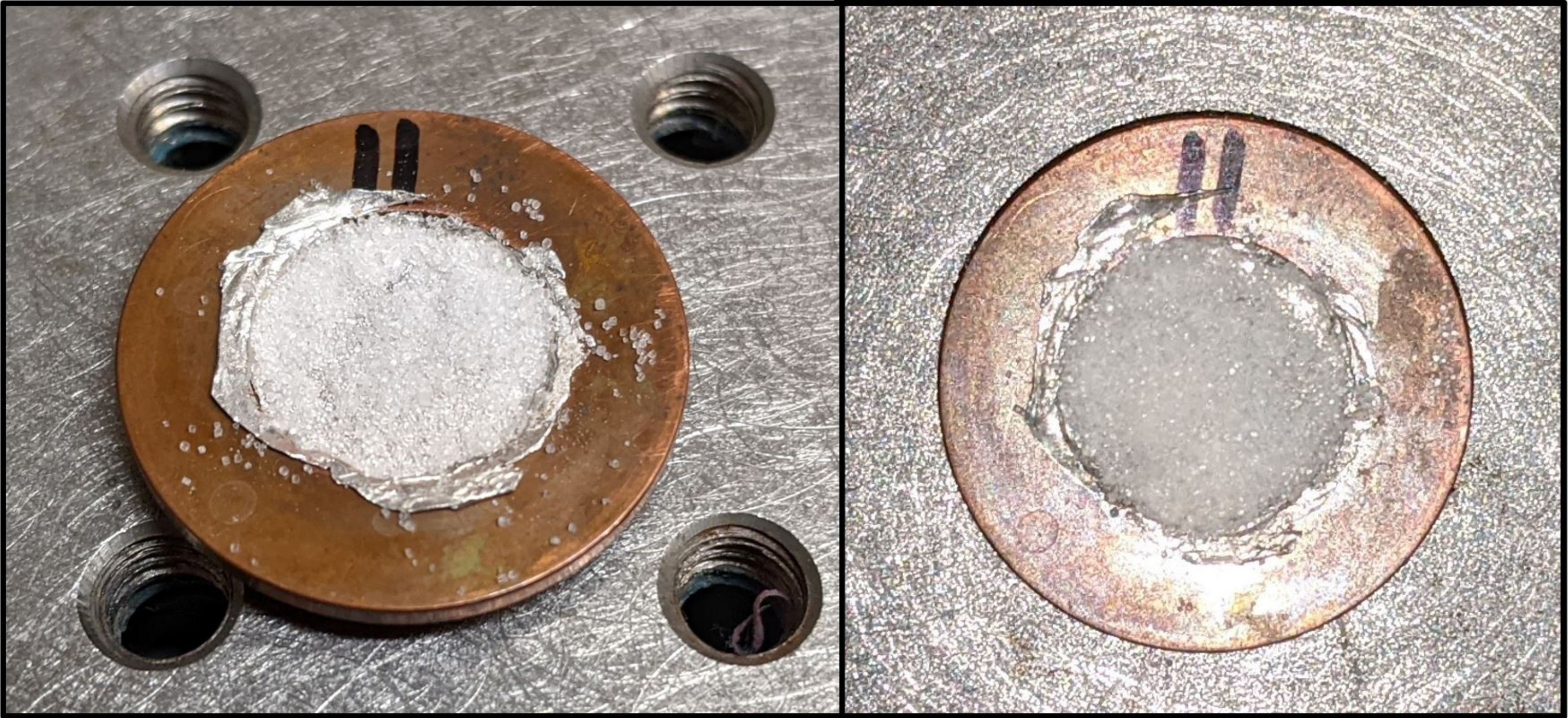}
    
    \caption{The sodium chloride sample after pressing into indium foil and 
    pouring off non-attached grains. The surface area of the sample predominantly shows sodium chloride grains, though in the specular 
    image on the right reflection can be seen from some small regions of
    bare indium.}
\end{figure}

\begin{figure}
    \centering
    \plotone{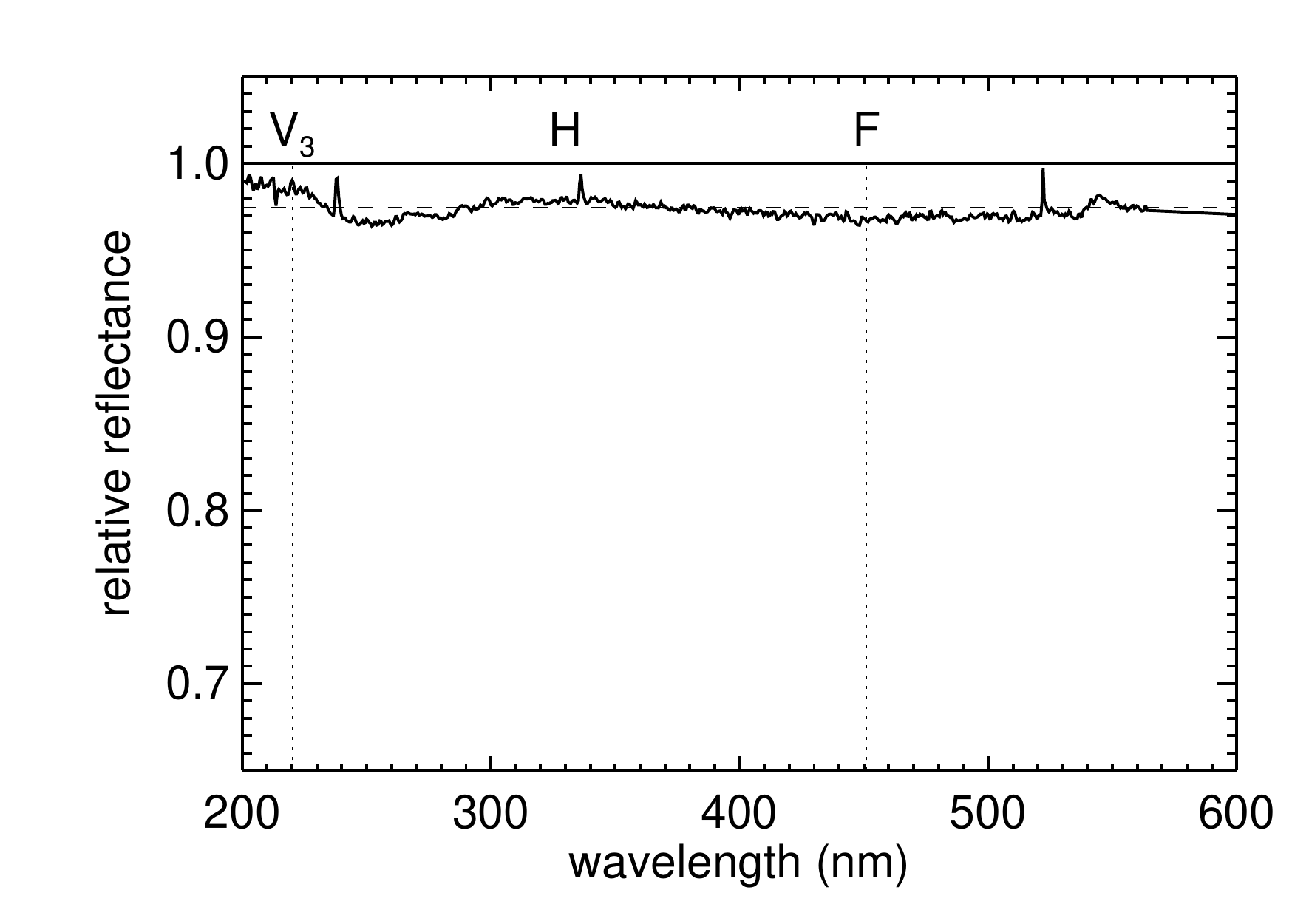}
    \caption{A spectrum of pure indium foil irradiated for three hours
    at identical conditions to the NaCl experiements divided by the spectrum
    of unirradiated indium (equivalent to the tourquoise-colored curves in Figure 1). Irradiation of  indium causes no features larger than $\pm 1.5$\% to 
    appear in the reflectance spectrum.}
\end{figure}


\begin{thebibliography}{}
\expandafter\ifx\csname natexlab\endcsname\relax\def\natexlab#1{#1}\fi
\providecommand{\url}[1]{\href{#1}{#1}}
\providecommand{\dodoi}[1]{doi:~\href{http://doi.org/#1}{\nolinkurl{#1}}}
\providecommand{\doeprint}[1]{\href{http://ascl.net/#1}{\nolinkurl{http://ascl.net/#1}}}
\providecommand{\doarXiv}[1]{\href{https://arxiv.org/abs/#1}{\nolinkurl{https://arxiv.org/abs/#1}}}

\bibitem[{Alexander \& Schneider(1950)}]{V3_1949}
Alexander, J., \& Schneider, E. 1950, Nature, 164, 653

\bibitem[{{Brown} \& {Hand}(2013)}]{2013AJ....145..110B}
{Brown}, M.~E., \& {Hand}, K.~P. 2013, \aj, 145, 110,
  \dodoi{10.1088/0004-6256/145/4/110}

\bibitem[{{Carlson} {et~al.}(1999){Carlson}, {Johnson}, \&
  {Anderson}}]{1999Sci...286...97C}
{Carlson}, R.~W., {Johnson}, R.~E., \& {Anderson}, M.~S. 1999, Science, 286,
  97, \dodoi{10.1126/science.286.5437.97}

\bibitem[{Casler {et~al.}(1950)Casler, Pringsheim, \& Yuster}]{V3}
Casler, R., Pringsheim, P., \& Yuster, P. 1950, J. Chem. Phys., 18, 1564

\bibitem[{{Dalton} {et~al.}(2012){Dalton}, {Shirley}, \&
  {Kamp}}]{2012JGRE..117.3003D}
{Dalton}, J.~B., I., {Shirley}, J.~H., \& {Kamp}, L.~W. 2012, Journal of
  Geophysical Research (Planets), 117, E03003, \dodoi{10.1029/2011JE003909}

\bibitem[{{Dalton} {et~al.}(2005){Dalton}, {Prieto-Ballesteros}, {Kargel},
  {Jamieson}, {Jolivet}, \& {Quinn}}]{2005Icar..177..472D}
{Dalton}, J.~B., {Prieto-Ballesteros}, O., {Kargel}, J.~S., {et~al.} 2005,
  \icarus, 177, 472, \dodoi{10.1016/j.icarus.2005.02.023}

\bibitem[{{Denman} {et~al.}(2022){Denman}, {Trumbo}, \&
  {Brown}}]{denman_nacl_2022}
{Denman}, W.~T., {Trumbo}, S.~K., \& {Brown}, M.~E. 2022, PSJ, in press

\bibitem[{{Fischer} {et~al.}(2015){Fischer}, {Brown}, \&
  {Hand}}]{2015AJ....150..164F}
{Fischer}, P.~D., {Brown}, M.~E., \& {Hand}, K.~P. 2015, \aj, 150, 164,
  \dodoi{10.1088/0004-6256/150/5/164}

\bibitem[{{Hand} \& {Carlson}(2015)}]{2015GeoRL..42.3174H}
{Hand}, K.~P., \& {Carlson}, R.~W. 2015, \grl, 42, 3174,
  \dodoi{10.1002/2015GL063559}

\bibitem[{{Johnson} {et~al.}(2019){Johnson}, {Hodyss}, {Vu}, \&
  {Choukroun}}]{2019Icar..321..857J}
{Johnson}, P.~V., {Hodyss}, R., {Vu}, T.~H., \& {Choukroun}, M. 2019, \icarus,
  321, 857, \dodoi{10.1016/j.icarus.2018.12.009}

\bibitem[{{Lellouch} {et~al.}(2003){Lellouch}, {Paubert}, {Moses}, {Schneider},
  \& {Strobel}}]{2003Natur.421...45L}
{Lellouch}, E., {Paubert}, G., {Moses}, J.~I., {Schneider}, N.~M., \&
  {Strobel}, D.~F. 2003, \nat, 421, 45, \dodoi{10.1038/nature01292}

\bibitem[{{Ligier} {et~al.}(2016){Ligier}, {Poulet}, {Carter}, {Brunetto}, \&
  {Gourgeot}}]{2016AJ....151..163L}
{Ligier}, N., {Poulet}, F., {Carter}, J., {Brunetto}, R., \& {Gourgeot}, F.
  2016, \aj, 151, 163, \dodoi{10.3847/0004-6256/151/6/163}

\bibitem[{{McCord} {et~al.}(1998){McCord}, {Hansen}, {Fanale}, {Carlson},
  {Matson}, {Johnson}, {Smythe}, {Crowley}, {Martin}, {Ocampo}, {Hibbitts}, \&
  {Granahan}}]{1998Sci...280.1242M}
{McCord}, T.~B., {Hansen}, G.~B., {Fanale}, F.~P., {et~al.} 1998, Science, 280,
  1242, \dodoi{10.1126/science.280.5367.1242}

\bibitem[{{McCord} {et~al.}(1999){McCord}, {Hansen}, {Matson}, {Jonhson},
  {Crowley}, {Fanale}, {Carlson}, {Smythe}, {Martin}, {Hibbitts}, {Granahan},
  \& {Ocampo}}]{1999JGR...10411827M}
{McCord}, T.~B., {Hansen}, G.~B., {Matson}, D.~L., {et~al.} 1999, \jgr, 104,
  11827, \dodoi{10.1029/1999JE900005}

\bibitem[{{Poston} {et~al.}(2017){Poston}, {Carlson}, \&
  {Hand}}]{2017JGRE..122.2644P}
{Poston}, M.~J., {Carlson}, R.~W., \& {Hand}, K.~P. 2017, Journal of
  Geophysical Research (Planets), 122, 2644, \dodoi{10.1002/2017JE005429}

\bibitem[{Schwartz {et~al.}(2008)Schwartz, Volkov, Sorokin, Trautmann, Voss,
  Neumann, \& Lang}]{PhysRevB.78.024120}
Schwartz, K., Volkov, A.~E., Sorokin, M.~V., {et~al.} 2008, Phys. Rev. B, 78,
  024120, \dodoi{10.1103/PhysRevB.78.024120}

\bibitem[{Seitz(1946)}]{RevModPhys.18.384}
Seitz, F. 1946, Rev. Mod. Phys., 18, 384, \dodoi{10.1103/RevModPhys.18.384}

\bibitem[{{Shirley} {et~al.}(2010){Shirley}, {Dalton}, {Prockter}, \&
  {Kamp}}]{2010Icar..210..358S}
{Shirley}, J.~H., {Dalton}, J.~B., {Prockter}, L.~M., \& {Kamp}, L.~W. 2010,
  \icarus, 210, 358, \dodoi{10.1016/j.icarus.2010.06.018}

\bibitem[{{Trumbo} {et~al.}(2019){Trumbo}, {Brown}, \&
  {Hand}}]{2019SciA....5.7123T}
{Trumbo}, S.~K., {Brown}, M.~E., \& {Hand}, K.~P. 2019, Science Advances, 5,
  aaw7123, \dodoi{10.1126/sciadv.aaw7123}

\bibitem[{{Trumbo} {et~al.}(2022){Trumbo}, {Becker}, {Brown}, {Denman},
  {Molyneux}, {Hendrix}, {Retherford}, , {Roth}, \&
  {Alday}}]{trumbo_HSTUV_2022}
{Trumbo}, S.~K., {Becker}, T.~M., {Brown}, M.~E., {et~al.} 2022, PSJ, in press

\bibitem[{Weidner \& Hsia(1981)}]{PTFE}
Weidner, V., \& Hsia, J. 1981, Journal of the Optical Society of America, 71,
  856

\end{thebibliography}
\end{document}